# Using singular value decomposition in a convolutional neural network to improve brain tumor segmentation accuracy

Pegah Ahadian, Maryam Babaei, Kourosh Parand

Department of Data and Computer Sciences, Shahid Beheshti University, Tehran, Iran

## ABSTRACT

*A brain tumor consists of cells showing abnormal brain growth. The area of the brain tumor significantly affects choosing the type of treatment and following the course of the disease during the treatment. At the same time, pictures of Brain MRIs are accompanied by noise. Eliminating existing noises can significantly impact the better segmentation and diagnosis of brain tumors. In this work, we have tried using the analysis of eigenvalues. We have used the MSVD algorithm, reducing the image noise and then using the deep neural network to segment the tumor in the images. The proposed method's accuracy was increased by 2.4% compared to using the original images. With Using the MSVD method, convergence speed has also increased, showing the proposed method's effectiveness.*

## KEYWORDS

*Artificial Neural Network, SV, Brain tumor, Brain MR, Segmentation*

## 1. INTRODUCTION

A brain tumor is an abnormal mass in the brain that can be cancerous (malignant) or non-cancerous (benign). Such as the type, location, size, and manner of tumor expansion that affect the degree of the threat of a tumor. Diagnosis of brain tumors due to size, form, and appearance in the brain is a complex task. Today, with the advancement of image segmentation algorithms, it is possible to diagnose a brain tumor. It is also provided automatically. These methods make it possible to store the state of tumor growth during treatment and reduce errors in human actions. However, due to the diversity of tumors, achieving high accuracy in tumor diagnosis is challenging. From diagnosis methods, a brain tumor can be detected by color histogram method, MRI image segmentation methods, and adaptive neuro-fuzzy inference system method. The artificial neural network was one of the available methods for detecting brain tumors in MRI images. The convolutional neural network is one of the deep methods of machine learning which is used due to its high efficiency. It has become prevalent to extract features and categorize images, which has a considerable image clarification impact. This way, the algorithm performance is improved by targeting the noises in the image [1]. The most critical removal methods are the noise of the images can be referred to as average filters, Wiener filters, non-local averages, and singular values decomposition. MSVD, which is obtained using the singular value analysis method, in addition to removing the noises of images, also preserves the quality of the images. This paper has tried firstly by choosing the appropriate number of individual values in the analysis of individual values of brain MRI images, to reduce the volume of existing noises in the images, and then by using a convolutional neural network method to detect brain tumors in the images.

## 2. METHOD

### 2.1. Singular Value Decomposition

The decomposition of singular values plays a crucial role in various fields of numerical linear algebra. Due to the properties and efficiency of this decomposition, many modern algorithms and methods are based on singular value decomposition. Singular Value Decomposition of an

[M x N] matrix A is obtained from the matrix equation:

$$A = USV^T$$

Where U is an orthonormal matrix of size [M x M] and V is an orthonormal matrix of size [N x N], and $V^T$ is transpose of V. S is a diagonal matrix (of singular values) of size [M x N] with positive values on the [M x M] in the following form:

$$S = \begin{bmatrix} \partial_1 & & \\ & \ddots & \\ & & \partial_n \end{bmatrix}$$

and

$$\partial_1 \geq \partial_2 \geq \partial_3 \geq \cdots \geq \partial_n$$

The elements in S are square roots of eigenvalues of $A^T A$. Moreover, the diagonal elements in S are sorted in descending order [2]. The SVD bases have frequency interpretations for images. As S(1,1) has the greatest value, its significance in describing image features is the highest. Values of the diagonal elements of S decrease and so also their significance in describing the image features.
One of the applications of this analysis is to use this method in image processing for Image compression is the removal of noise and scratches in a photo, etc. [3]. Usually, matrices corresponding to images have a large number of values that are small singulars. Suppose there is (n − k) a small singular value of matrix A that can be neglected. Then the matrix $\partial_1 u_1 \vartheta^T{}_1 + \partial_2 u_2 \vartheta^T{}_2 + \ldots + \partial_k u_k \vartheta^T{}_k$ is an excellent approximation of A, and such an approximation can be sufficient in applications [4]. Fig. 1 shows the K-SVD method's effect on MRI images by choosing different K values.
As part of this study, we used a multiresolution SVD for image denoising. Multiresolution singular value decomposition (MSVD) method is similar to the multiresolution analysis (MRA) employed in the wavelet domain [5]. The noisy image is decomposed into four sub-bands of varying frequency in the first level of MSVD decomposition. The low-frequency sub-band represents most of the signal. As noise is generally of high frequency, it is present primarily in the detailed high-frequency sub-bands. The present noise in detailed sub-bands is removed using suitable thresholding techniques. Synthesis of the approximation sub band and threshold detailed sub-band is done using inverse multiresolution SVD (IMSVD) to get the denoised image. MSVD algorithm is given in the following semi-code:

*Input Image A*

```
A1 = Reshape(A. 4.16384)
[U S V] = svd(A1)
T = U^T A1
Y.LL = Reshape(T(1.:). 128.128)
Y.LH = Reshape(T(2.:). 128.128)
Y.HL = Reshape(T(3.:). 128.128)
Y.HH = Reshape(T(4.:). 128.128)
T1(1.:) = Reshape( (Y.LL). 1.16384)
T1(2.:) = Reshape ((Y.LH). 1.16384)
T1(3.:) = Reshape( (Y.HL). 1.16384)
T1(4.:) = Reshape ((Y.HH). 1.16384)
A2 = UT1 A2 is (4.16384)
A = Reshape(A2 .256.256)
Output A
```

algorithm 1: MSVD algorithm

The Y.LL band contains low-frequency details. The other three contain high-frequency (horizontal, vertical, and diagonal) details.

### 2.2. Convolutional Neural Network

Among machine learning methods, an artificial neural network is a popular method that simulates the neural network mechanisms of biological organisms. It consists of a network of nodes called neurons and weighted edges. In this system, the inputs flow in the network, and a series of outputs are produced. Then the outputs are compared with valid data, and the paths that lead to the detection of correct cases will be strengthened and the network paths with the wrong detection corrected. The network can finally achieve perfect accuracy and the desired task by repeating this process. One of the essential steps in the development of artificial neural networks is the design of the network architecture, which greatly affects network performance. In designing the architecture of artificial neural networks, factors such as the number of hidden layers, the number of neurons in each layer, transformation functions, and training algorithms must be determined.

One of the common architectures is called a convolutional neural network (CNN). Convolutional neural networks are a class of deep neural networks that are usually used to perform image or speech analysis in machine learning. The fundamental motivation for these networks is that the recognition function of the visual part of the cat's brain was found, which seemed to stimulate specific neurons in certain parts of the visual field. This rule was more widely used to design the architecture of these networks. In convolutional neural network architecture, each layer of the network is 3-dimensional, with a spatial range and depth corresponding to the number of features. Historically convolutional neural networks are the most successful type of neural network. These networks are widely used for image recognition, object localization, and even text processing, and the performance of these networks has recently surpassed humans in the problem of image classification.

One of the common architectures is called a convolutional neural network (CNN). Convolutional neural networks are a class of deep neural networks that are usually used to perform image or speech analysis in machine learning. The fundamental motivation for these networks is that the recognition function of the visual part of the cat's brain was found, which seemed to stimulate specific neurons in certain parts of the visual field. This rule was more widely used to design the architecture of these networks. In convolutional neural network architecture, each layer of the network is 3-dimensional, with a spatial range and depth corresponding to the number of features. Historically convolutional neural networks are the most successful type of neural network. These networks are widely used for image recognition, object localization, and even text processing, and the performance of these networks has recently surpassed humans in the problem of image classification.

### 2.3. U-net in tumor segmentation

In this work, we first tried to reduce the noises in the images by using the MSVD method and detecting brain tumors with image segmentation. Image segmentation involves partitioning an input image into different segments with a strong correlation with the region of interest in the given image. The field of medical image analysis is growing, and the segmentation of the organs, diseases, or abnormalities in medical images has become demanding [6]. Medical image segmentation aims to represent a given input image in a meaningful form to study the volume of tissue in order to measure the tumor size and help in deciding the dose of medicine, planning of treatment prior to applying radiation therapy, or calculating the radiation dose. Recently, deep neural models have shown application in various image segmentation tasks. We used the U-Net, a convolutional network for Biomedical Image Segmentation. The U-NET was developed by Olaf Ranneberger et al. for Biomedical Image Segmentation [7]. The architecture contains two paths. The encoder path is used to capture the context of the image. The encoder is just a traditional stack of convolutional and max pooling layers. The decoder is the symmetric expanding path that is used to enable precise localization using transposed convolutions. This network only contains convolutional layers and does not contain any dense layer because of which it can accept images of any size. The neural network architecture of the U-Net that we use in this paper is given in figure 2.

In general, the proposed structure includes a part to reduce image noise, and another part is a U-Net neural network for brain tumor segmentation.

### 2.4. Experimental results

To check the effectiveness of the proposed method, we have considered a brain MRI dataset, including 3900 images with a size of 256,256 pixels. The images were obtained from The Cancer Imaging Archive (TCIA). After that, the images were improved using the MSVD method. Then, diagnosed images are used as input to the U-Net neural network. With the Adam optimizer, the learning rate is set to 1e-4. We have also employed data augmentation methods to improve the robustness of the model, including scaling, shifting, flipping, and horizontal flipping. The model is trained using 70 epochs. The U-Net architecture of the proposed method is illustrated in Fig. 2. In Fig. 3, we present the network's outputs and an MRI image with the prediction mask. We have considered three evaluation metrics to compare the results: binary accuracy, Intersection-Over-Union (IOU), and Dice Coefficient, which are defined as follows:

$$\text{Binary accy} = \frac{TP + TN}{TP + TN + FP + FN}$$

$$IOU = \frac{Area\ of\ Overlap}{Area\ of\ Union}$$

$$Dice = \frac{2 \times Area\ of\ Overlap}{Total\ number\ of\ pixels}$$

A comparison of final evaluation metrics is displayed in table 1. Furthermore, we can see that using denoised images causes faster network convergence in Fig. 4.

Table 1. comparison of final evaluation metrics

| *data* | *Binary accuracy* | *IOU* | *Dice-coef* |
|---|---|---|---|
| Orginal image | 0.9980 | 0.7984 | 0.8865 |
| Denoised image with MSVD | 0.9982 | 0.8116 | 0.8946 |

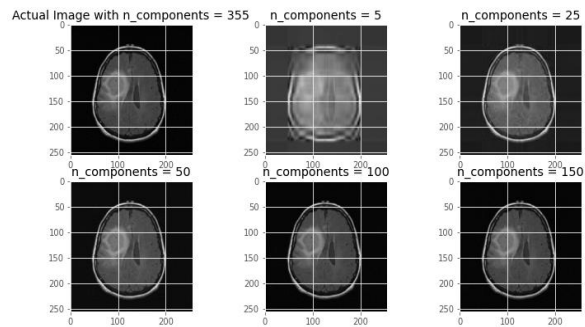

Figure 1. The effect of using the K-SVD method on MRI images by choosing different K values

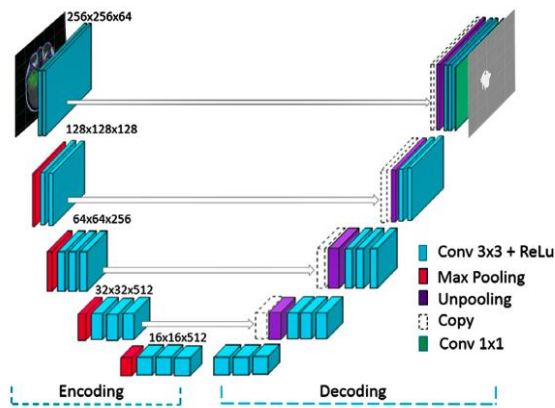

Figure 2. U-Net architecture related to mentioned method

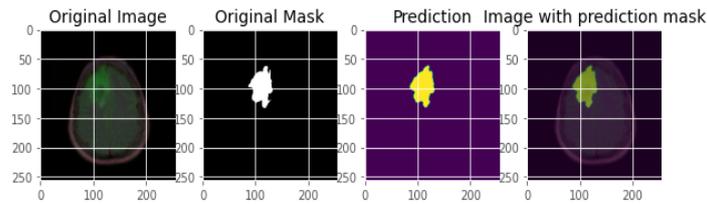

Figure 3. comparing the network output with the original mask

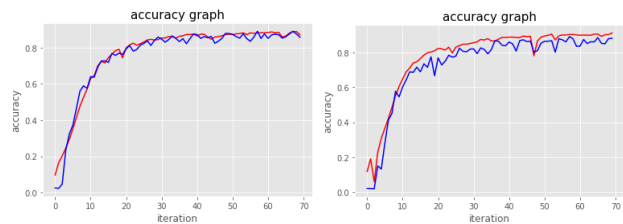

Figure 4. Comparing the accuracy of U-Net with (a) denoised data and (b) initial data

## 3. CONCLUSIONS

One of the challenges in processing and diagnosing diseases according to MRI and CT scan images is reducing the image noise, which can be achieved by analyzing individual values. This research investigates the effect of using the MSVD method on speed and accuracy detection in a convolutional neural network. The results show that using a singular value experiment to destroy the noise and denoise the images has reduced the execution time of each epoch of neural network learning. Besides, the speed of convergence of the network has increased significantly.

**Authors**

Pegah Ahadian

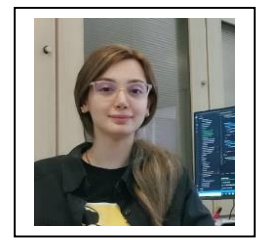